\documentclass[onecolumn,showpacs,preprintnumbers,amsmath,amssymb,10pt]{revtex4}

\usepackage{epsf}
\usepackage{graphicx}  
\usepackage{dcolumn}   
\usepackage{bm}        

\newcommand{\be}{\begin{equation}}
\newcommand{\en}{\end{equation}}
\newcommand{\bea}{\begin{eqnarray}}
\newcommand{\ena}{\end{eqnarray}}

\begin{document}

\preprint{SUCA/008-2010}

\title{De Sitter ground state of scalar-tensor gravity and its primordial perturbation}

\author{Hongsheng Zhang\footnote{Electronic address: hongsheng@shnu.edu.cn} }
\affiliation{Shanghai United Center for Astrophysics (SUCA),
 Shanghai Normal University, 100 Guilin Road, Shanghai 200234,
 P.R.China}
\author{Xin-Zhou Li \footnote{Electronic address: kychz@shnu.edu.cn} }
 \affiliation{Shanghai United Center for Astrophysics (SUCA), Shanghai Normal
University, 100 Guilin Road, Shanghai 200234, P.R.China}

\begin{abstract}
   Scalar-tensor gravity is one of the most competitive gravity theory to Einstein's relativity. We reconstruct the exact de Sitter solution in  scalar-tensor gravity, in which the non-minimal coupling scalar is rolling along the potential. This solution may have some relation to the early inflation and present acceleration of the universe.
    We investigated its primordial quantum perturbation around the
   adiabatic vacuum.   We put forward for the first time that exact de Sitter generates non-exactly scale invariant perturbations.
     In the conformal coupling case, this model predicts that the tensor mode of the perturbation
     (gravity wave) is strongly depressed.
\end{abstract}


\maketitle

 \section{Introduction} The scalar-tensor gravity (STG) is one of the most influential competitors to
  Einstein's general relativity. Contrarily to common idea, it is compulsive in several cases that a non-minimal coupled scalar is involved
  in the propagation of gravity. The most competitive reason comes from quantum arguments. When quantum correction is considered, or renormalizability of the
  scalar field in curved space is required, the product term of $\phi^2 R$ is inevitable \cite{qcs}.  We always meet such terms when we make
  dimensional reductions from higher dimensional theories, such as string/M.

 For any physical theories, especially for the non-linear theories, exact mathematical solutions is inevitably to be one of the central topic ever since
 the set up of a theory. For a non-linear theory,  generally the algebraic sum of two independent solutions is not a solution:
 the properties of approximate solution may be far away from the real exact solution. The
 STG is also highly non-linear as Einstein's theory, even more non-linear than it. A few exact solutions of STG cosmology has been found \cite{exact1, exact2}.  However, to
 our knowledge, technically a most significant solution, de Sitter solution, is still absent. Since both the early universe (inflationary stage) and the late-time universe (present cosmic acceleration) are quasi-de Sitter phases, we expect such a solution may have some relation to the realistic universe.
 In this work we construct such a solution, in which the scalar is rolling down along a fine-tuned multinomial power-law potential. Furthermore, the energy density and pressure keep constant, which seems impossible for the minimal coupling case.

 Some original ideas of STG can be traced back to Mach's principle and Dirac's proposal, which suggests that the gravity constant may be a variable. Brans and Dicke developed a mathematical formalism to realize this idea, in which a non-minimally coupled scalar field was introduced \cite{br}. Advancing a little step along this direction, ie, to introduce a potential for the non-minimally coupled scalar, we reaches the simplest STG theory. Immediately after the inflation was proposed in general relativity, it was studied in frame STG \cite{1stinf}. Pioneering works of the effect of the non-minimal coupling term on CMB is proposed in \cite{bo}. The relation between the GUT scale and the coupling constant is explored in \cite{salo2}. It is suggested that some problems of the chaotic inflation alleviate via a non-minimal coupling term \cite{fu}. Under some reasonable situation, the coupling constant should be very small or negative in inflationary scenarios \cite{mada}. The inflation model of the Standard Model Higgs boson with non-minimal coupling to the gravity is re-examined recently \cite{barvin}.  Inflation in STG with  various potentials were also  systematically examined in \cite{fara}. The inflation model in frame of STG is also studied in \cite{noninf}. The inflation model can solve some problems in the original big model, and more significantly, it presents an elegant mechanism to generate the primordial perturbation as the seeds of structure formation.
    The perturbation spectrum in frame of scalar tensor theory with a $\phi^4$ potential is checked in \cite{sasaki}. Second-order matter density perturbations and skewness in STG is investigated in \cite{second}. The non-linear perturbation in STG cosmology is explored in \cite{perro}. The researches on perturbation growth of matter for STG dark energy are proposed in \cite{pergrow}.    A gauge invariant approach on perturbations of STG cosmology is suggested in \cite{inva}. The cosmological perturbation theory has been made based on generalized gravity theories including string corrections and tachyon \cite{hn}.  We extend this study to
 STG, and apply it to generalized slow-roll inflation and its consequent power spectra.  We put forward for the first time that exact de Sitter generates non-exactly scale invariant perturbations.     This model predicts that the tensor mode of the perturbation is strongly depressed in the case of conformal coupling.

  In the next section we shall study the de Sitter solution for the STG.  Next comes the primordial perturbation of the solution. We present our conclusion
     and some discussions at the end of this article.

 \section{De~ Sitter~ state ~for~ STG}  We start from the action of STG,
   \be
   S= \int_{M} d^4 x\sqrt{-g} \left[
   {1\over 2\kappa}R-\frac{\xi}{2}\phi^2 R+L_{\rm scalar}(\phi)+L_{\rm matter}\right]+{1\over \kappa}\int_{\partial M} d^3 x\sqrt{-h} K.
       \label{actionSTG}
       \en
       Here $\kappa$ is the gravity constant, $M$ represents the spacetime manifold, $K$ denotes the extrinsic curvature of its boundary,
       $R$ is the Ricci scalar, $\xi$ is a constant, $h$, $g$ stand for the determinants of the 4-metric $g_{\mu\nu}$ and its induced 3-metric $h_{\mu\nu}$ on the boundary, respectively.
       $L_{\rm scalar}$ and $L_{\rm matter}$ are the Lagrangians of the non-minimally coupled scalar and other minimally coupled matters to gravity, respectively.
  $L_{\rm scalar}$ takes the same form of an ordinary scalar,
  \be
   L_{\rm scalar}=-\frac{1}{2}g^{\mu\nu}\nabla_{\mu}\phi\nabla_{\nu}\phi-V(\phi).
   \label{actionS}
   \en
   By making variation of (\ref{actionSTG}) with respect to $\phi$, one obtains the equation of motion of the scalar,
   \be
   g^{\mu\nu}\nabla_{\mu}\nabla_{\nu}\phi-\xi R\phi-\frac{dV}{d\phi}=0.
   \label{eom}
   \en
  To the equation of motion for $\phi$, the non-minimal coupling term  $\xi\phi^2R/2$ is just equivalent to an extra potential.
  Variation of (\ref{actionSTG}) with respect
  to $g_{\mu\nu}$ yields the field equation,

  \be
  (1-\kappa \xi \phi^2)G^{\mu\nu}=\kappa\left[T^{\mu\nu}(\phi)+T^{\mu\nu}({\rm matter})\right],
  \label{FE}
  \en
 where $G^{\mu\nu}$ denotes Einstein tensor, $T^{\mu\nu}({\rm matter})$ labels the energy-momentum tensor corresponding to
 $L_{\rm matter}$ in (\ref{actionSTG}), and $T^{\mu\nu}(\phi)$ takes the following form,

 \be
 T^{\mu\nu}(\phi)=\nabla^{\mu}\phi\nabla^{\nu}\phi-\frac{1}{2}g^{\mu\nu}g^{\alpha\beta}\nabla_{\alpha}\phi\nabla_{\beta}\phi-
 Vg^{\mu\nu}+\xi\left[g^{\mu\nu}g^{\alpha\beta}\nabla_{\alpha}\nabla_{\beta}(\phi^2)-\nabla^{\mu}\nabla^{\nu}(\phi^2)\right].
 \label{emnmt}
 \en
 For a detailed deduction of (\ref{FE}), see the appendix C in \cite{book}.  It deserves to note that one does not need to introduce new boundary
 term other than $K$ in this derivation, since one can repeatedly apply integration by parts to remove all the derivation terms of $g_{\mu\nu}$
 yielded by $\phi^2R$ on the boundary. If we define an effective gravity constant $\kappa_{\rm eff}$,
 \be
 \kappa_{\rm eff}=(1-\kappa \xi \phi^2)^{-1}\kappa,
 \en
 then the field equation (\ref{FE}) reduces to  Einstein form with a variable gravity ``constant". That is the original idea of Brans-Dicke proposal.
 When $\phi\to 0$, STG degenerates to standard general relativity. An hence both Minkowski and (anti-)de Sitter are permitted. But it is only a trivial case of
 STG. How about a non-zero $\phi$?

 Now we try to find the solution of STG with maximally symmetric space. A maximally symmetric space (in a proper chart) can be defined as an FRW universe with constant Hubble parameter.
 In an FRW universe, the field equation becomes Friedmann equations,
 \be
 (1-\kappa \xi \phi^2)\left(H^2+\frac{k}{a^2}\right)=\frac{\kappa}{3}\rho,
 \label{fried1}
 \en
 \be
 (1-\kappa \xi \phi^2)\frac{\ddot{a}}{a}=-\frac{\kappa}{6}(\rho+3p),
 \label{fried2}
 \en
 where $a$ is the scale factor, the term $(1-\kappa \xi \phi^2)$ highlights that $\phi$ is involved in gravity interaction, $\rho$ and $p$ denote
 the total density and pressure,
 \be
 \rho=\frac{1}{2}\dot{\phi}^2+V+6\xi \phi \dot{\phi}H+\rho_{\rm matter},
 \en
 \be
 p=\frac{1-4\xi}{2}\dot{\phi}^2-V-2\xi \phi \ddot{\phi}-4\xi \phi \dot{\phi}H+p_{\rm matter}.
  \en
 Here matter labels all the matters other than $\phi$. To find the ground state of STG, we  consider the its vacuum solution without any other
 matter fields other $\phi$. In the next paper we shall study a dust fluctuation on this background \cite{self2}.
 In the studies of dark energy, a model-independent method for estimating the form of the potential
 $V$  of the scalar field  which drives the cosmic acceleration from observation data is developed in
 \cite{star2}. This method is dubbed reconstruction. And then, the reconstruction of scalar
 tensor cosmology is developed in \cite{star3}. For a review of the
 reconstruction method in cosmology, see \cite{starreview}. A very
 recent review of the reconstruction by data of modified gravity
 theory is presented \cite{odin}.

  We reconstruct the following solution of (\ref{fried1}) and
 (\ref{fried2}) by setting $\rho_{\rm matter}=0$ and $p_{\rm matter}=0$,
 \be
 \phi=c_2\left[e^{2c_1b\xi}-e^{bt}(4\xi-1)\right]^{2\xi \over {4\xi-1}},
 \label{phis}
 \en
 \be
 V=\frac{3b^2}{\kappa}-\frac{\xi b^2}{\phi^2 (1-4\xi)^2}\left[(3-34\xi+96\xi^2)\phi^4+
 8\xi c_2^{2-\frac{1}{2\xi}}e^{2c_1b\xi}(6\xi-1)\phi^{2+\frac{1}{2\xi}}+2\xi c_2^{4-\frac{1}{\xi}}e^{4c_1b\xi}\phi^{\frac{1}{\xi}}\right],
 \label{potential}
 \en

 \be
 \rho=-p=\frac{3b^2}{\kappa}(1-\kappa \xi \phi^2),
 \label{rhop}
 \en

 \be
 a=c_3e^{bt},
 \label{scale}
 \en
  \be
  k=0,
  \label{mo}
  \en

 where $c_1,~c_2,~c_3$ are integration constants. It is easy to see that the above set-up describes a de Sitter
 space.
 Now we make some notes on this solution. First $b$ is the energy scale of the universe. It should be noted that we reconstruct the exact de Sitter
 space in STG, the energy scale of which is arbitrary. The potential
 of the approximate de Sitter (quasi-de Sitter in slow roll
 inflation) constructed by data is narrow \cite{peri}. It is clear that the solution becomes
 a Minkowskian one when $b=0$. In an expanding universe $b>0$. $c_2$ is an constant factor in $\phi$ and $c_2=0$ yields an ordinary de Sitter solution in general relativity.
 $c_3$ is a constant in the scale factor and does not appear in other physical quantities.
 There is another important requirement on $\phi$: we do not need a complex field, which we called real condition. For the case  $\xi\le 1/4$, thus including the conformal coupling case,
 real condition has no constraint on the parameters. For the case $\xi>1/4$, $\phi$ may be divergent and even be a imaginary number when $t$ is big enough. In fact $\phi$ may be complex if $m$ is an even number, where
 the irreducible fraction $\frac{n}{m}={\xi \over {4\xi-1}}$. So in the case $\xi>1/4$ and $m$ is even, we define
 \be
 \phi=c_2\left[e^{2c_1b\xi}-e^{bt}(4\xi-1)\right]^{2\xi \over {4\xi-1}},~~~~~~~ {\rm for}~~~~~~~e^{2c_1b\xi}-e^{bt}(4\xi-1)>0,
 \en

  \be
 \phi=c_2\left[-e^{2c_1b\xi}+e^{bt}(4\xi-1)\right]^{2\xi \over {4\xi-1}},~~~~~~~ {\rm for}~~~~~~~e^{2c_1b\xi}-e^{bt}(4\xi-1)<0.
 \en
 All of the other quantities in this set-up are not changed. Physically, at $e^{2c_1b\xi}-e^{bt}(4\xi-1)=0$, the field $\phi$ rotates an
 unobservable global angle in the inner space. Since $\xi>1/4$, the exponent $2\xi \over {4\xi-1}>0$, thus the phase of $\phi$ is arbitrary  at  $e^{2c_1b\xi}-e^{bt}(4\xi-1)=0$ where $\phi=0$.

 Interestingly, though the density and pressure (\ref{rhop}) are not constant in the evolution of the spacetime, they keep
 cunning counteraction with the extra factor in the modified Friedmann equation (\ref{fried1}) and
 (\ref{fried2}). And hence the Hubble parameter can be a constant with a rolling scalar along the potential (\ref{potential}).
 The conformal coupling case is the most important case in the non-minimal coupling theory. We thus prove that for STG de Sitter
 space does not mean a vacuum dominated space. A rolling scalar also can lead to an exact de Sitter.  For the conformal coupling case $\xi=1/6$,
  the potential $V$ degenerates to an extraordinary simple form,
  \be
  V= \frac{3b^2}{\kappa}-\frac{e^{2c_1b/3}b^2}{2c_2^2}\phi^4,
  \label{vsimple}
  \en
 while the other quantities in the above set-up become,
 \be
 \phi=c_2\left(e^{c_1b/3}+\frac{1}{3}e^{bt}\right)^{-1},
 \label{phisimple}
 \en

 \be
 \rho=-p=\frac{3b^2}{\kappa}(1-\frac{1}{6}\kappa  \phi^2),
 \label{rhosim}
 \en

 \be
 a=c_3e^{bt},
 \en
  \be
  k=0.
  \en
 For this conformal coupling case, we note that both $\phi$ and $a$ are exponential functions of $t$ when $c_1\to -\infty$,
  which never appears in general relativity. The potential (\ref{vsimple}) has no lower bound, and thus stability problem may appear, which will be studied in our next work.
   Here we only note that its dynamics is not very similar to an ordinary scalar since the Ricci scalar $R$ is involved in its dynamics. From (\ref{eom}), the effective mass comes not only from $V$, but also from $R$.

   Now we present a preliminary study of the stability of this de Sitter space of STG. A nice research to the stability of the inflation model in the Einstein gravity minimally coupled to a scalar via a Hamilton-Jacobi formulism has been developed in \cite{salo}. Here we apply their method to our model. The essential operation of the Hamilton-Jacobi method in inflation is to replace $\dot{\phi}$ with the Hubble parameter $H(\phi)$, and thus we get an equation with only variables $H(\phi)$ and $V(\phi)$. In the non-minimal coupling case, it is difficult to write $\dot{\phi}$ as an explicit function of $H(\phi)$ since both the equation of motion of $\phi$ (\ref{eom}) and the Friedmann equation (\ref{fried1}) are rather complicated. Under this situation, we explore the weak coupling limit as the first step. Differentiating with respect to $t$ at both sides of (\ref{fried1}), and then substituting to (\ref{eom}), we obtain
   \be
   H'=-\frac{\kappa}{2}\dot{\phi}+\kappa\xi\dot{\phi},
   \en
   in the weak coupling limit. Here a prime denotes differentiation with respect to $\phi$. It is clear that the term $\kappa\xi\dot{\phi}$ is the correction to the standard model \cite{salo}, which will recover when $\xi=0$. Considering a linear perturbation of $H(\phi)=H_0(\phi)+\delta H(\phi)$, we find the perturbation evolves as
   \be
   H\delta H=\frac{2\kappa}{3}H'\delta H'(-1+2\xi)^{-2}.
   \en
   The general solution of the above equation reads,
   \be
   \delta H(\phi)=\delta{\phi_i}\exp \left(\frac{3}{2\kappa}(-1+2\xi)^{2}\int_{\phi_i}^{\phi}\frac{H_0}{H_0'}d\phi\right),
   \en
  where $i$ labels some initial value of $\phi$.
  Since the integrand is negative definite, the linear perturbation will exponentially damp away,
  and hence this solution is stable against linear perturbations in the weak coupling
  limit. The pre-exponential of the about equation denotes the
  variation of the Hubble parameter.  We must be aware that the stability property of this solution may be
  very different in the strong coupling case, which deserves to explore further.
  \section{Einstein Frame}
  In the above discussions, we work in Jordan frame, which is a natural frame for STG. The explanation of observation results depends on  the ``ansats" of the frames\cite{sasa}. The
  frame which is more familiar to our experience is Einstein frame. Thus, it may be useful to see the form of our solution in the Einstein frame. Based on the form of
  action of STG (\ref{actionSTG}), we introduce a conformal transformation as follows,

  \be
  \bar{g_{\mu\nu}}=\Omega g_{\mu\nu},
  \en
  where
  \be
  \Omega=1-\kappa \xi \phi^2,
  \en
  where a bar implies a quantity in Einstein frame.
  As a result we have
  \be
  \bar{a}=\Omega^{1/2}~a,
  \label{bara}
  \en
  \be
  d\bar{t}^2=\Omega dt^2.
  \label{ttran}
  \en
  From (\ref{scale}) and (\ref{ttran}), we get the exact form of the scale factor in parametric form. However, since the representations are a little bit involved, the physical sense of
  this solution is still obscure. To clarify the physics of this solution, we calculate the deceleration parameter $\bar{q}$, which carries the effect of the geometric evolution of the universe.
  \be
  \bar{q}=-\frac{\ddot{\bar{a}}\bar{a}}{\dot{\bar{a}}^2},
  \en
  where a dot indicates derivative with respect to $\bar{t}$.

 After tedious but straitforwards calculations, we reach
 \bea
 \nonumber
 \bar{q}=\left\{\frac{\phi^2}{c_2^2}+\kappa \phi^4\xi^2[e^{4bc_1\xi}+2(1-6\xi)e^{b(t+2c_1\xi)}+(1-4\xi)(1-6\xi)e^{2bt}]
 \right. \\\left.-2\kappa\phi^2\xi[e^{4bc_1\xi}+2(1-5\xi)e^{b(t+2c_1\xi)}+(1-9\xi+24\xi^2)e^{2bt}]\right\}/\left\{4\xi e^{bt}+
 \kappa \xi \phi^2[e^{2bc_1\xi}+(1-6\xi)e^{bt}]-e^{bt}-e^{2bc_1\xi}\right\},
 \label{barq}
 \ena
 where $\phi$ is given by (\ref{phis}). Both $\bar{a}$ and $\bar{q}$ are functions of the Einstein time $t$, therefore $\bar{q}$ in (\ref{barq}) is an implicit function of $\bar{a}$ (\ref{bara}). But it is difficult to obtain the explicit form in the general case. We consider a special case of this solution, in which we set $\xi=1/6,~\kappa=c_2^{-2}, c_1\to -\infty$. In this case $\bar{q}$ can be written in an explicit function of $\bar{a}$,
 \be
 \bar{q}=\frac{6-2\bar{a}^2/c_3^2}{3+2\bar{a}^2/c_3^2}.
 \label{dece}
 \en
 Note that we did not set any special value for the parameter $b$, which is the energy scale of the de Sitter solution of STG. As the result, the $q$ in (\ref{dece}) can describe
 a high energy universe or a low one. It is clear when $\bar{a}\to \infty$ the universe becomes a de Sitter and when $\bar{a}\to 0$~~ $\bar{q}=2$, which equals a universe with stiff matter with the equation of state $w=1$. It is well-known that this stiff matter can be simulated by a free scalar.

 Further if we require this model describe our present evolution of the universe,  ie, $\bar{q}=-0.6$ when $\bar{a}=1$, we can determine $c_3=0.32$.  So our preliminary conclusion of this section is that in Einstein frame the de Sitter solution of STG can describe a universe which decelerates in the early time and accelerates in the late time. We hope this model has some relation with our realistic universe with dust matter.

  \section{Primordial ~spectra}
   From the above discussions, we see that this de Sitter solution is very special in that the scalar is dynamical rather than ``rests" on some
 position of the potential. A common lore says that a rolling scalar yields non-exact scale invariant perturbation, and an exact de Sitter
 generates exact scale invariant perturbation. The arguments are conflicting on the surface, which lead to a question: Whether is the perturbation scale invariant, or not? One must be curious the properties of the perturbation in this interesting solution.

 The quantum mechanism of primordial fluctuations in the generalized gravity including string corrections and tachyon has been discussed in detail in \cite{hn}. Now we extend this study to STG. Following that study, we define the
 following parameters to describe the generation of primordial perturbation,

 \be
 \epsilon_1=\frac{\dot{H}}{H^2},
 \en
  \be
 \epsilon_2=\frac{{\ddot{\phi}}}{H\dot{\phi}},
 \en
  \be
 \epsilon_3=-\frac{\xi\phi\dot{\phi}}{H(1/\kappa-\xi \phi^2)},
 \en
  \be
 \epsilon_4=\frac{\xi\phi\dot{\phi}(6\xi-1)}{[1/\kappa+\xi\phi^2(6\xi-1)]H}.
 \en
 In the above equations, we have made some calculations compared with the original definitions of $\epsilon_i, (i=1,2,3,4)$ in \cite{hn}.
 $\epsilon_{1}$ and  $\epsilon_{2}$ are directly inherited from general relativity, while $\epsilon_{3}$ and $\epsilon_{4}$ are new parameters for attached freedoms to STG.
 Under a slow-roll approximation, $|\epsilon_{i}|<<1$, we get the power spectrum of the adiabatic vacuum perturbation  ${\cal P}_{S}$ on large scale  \cite{hn},
 \be
  {\cal P}^{1/2}_{S} (k, \eta)
   = \left| {H \over 2 \pi z} \Big\{ 1 + \epsilon_1
       + \left( 2 \epsilon_1 - \epsilon_2 + \epsilon_3 - \epsilon_4 \right)
       \big[ \ln{(k|\beta|)} - 2 + \ln{2} + \gamma \big] \Big\}
       {} \right|,
 \en
 where
 \be
 z=\frac{\dot{\phi}}{H(1+\epsilon_3)}\sqrt{1+\frac{3\dot{F}^2}{2F\dot{\phi}^2}},
 \en
  \be
  \beta = - {1\over aH} {1\over 1 + \epsilon_1},
     \en
     \be
     F=\frac{1}{\kappa}-\xi\dot{\phi}^2,
     \en
     and $\gamma=0.577...$ is the Euler's constant.
   Then the spectral index of scalar perturbation $n_S$ reads,
 \be
 n_S-1=\frac{d\ln {\cal P}_S}{d\ln k}={2(2\epsilon_1-\epsilon_2+\epsilon_3-\epsilon_4)},
 \en
  and in STG, the consistency holds,
 \be
 r\triangleq \frac{\cal{P}_T}{\cal{P}_S}=2|n_T|,
 \en
 where $\cal{P}_T$ is the kernel of tensor perturbation mode in $\eta$-space ($\eta$: wave number).

 Thus we calculate the tensor perturbation index $n_T$,
 \be
 n_T={2(\epsilon_1-\epsilon_3)}.
 \en
 For our solution (\ref{phis})-(\ref{mo}), direct calculations give
 \be
  {\cal P}^{1/2}_{S}(k=aH)=\frac{be^{-bt}}{4\pi c_2\xi}A^{1+\frac{2\xi}{1-4\xi}}\left(1+\frac{6c_2^2\kappa \xi^2 A^{\frac{4\xi}{4\xi-1}}}{1-c_2^2\kappa \xi A^{\frac{4\xi}{4\xi-1}}}
  \right)^{-1/2},
 \label{ps}
 \en

 \be
 n_S-1=\frac{2(e^{2c_1b\xi}-2\xi e^{bt})}{A}+\frac{4c_2^2\kappa \xi^2 e^{bt}A^{\frac{1}{4\xi-1}}}{c_2^2\kappa \xi A^{\frac{4\xi}{4\xi-1}}-1}
 -\frac{4c_2^2\kappa \xi^2 e^{bt}(6\xi-1)A^{\frac{1}{4\xi-1}}}{1+c_2^2\kappa \xi (6\xi-1) A^{\frac{4\xi}{4\xi-1}}}
 ,
 \label{ns}
 \en

 \be
 n_T=\frac{4c_2^2\kappa\xi^2 e^{bt}A^{\frac{1}{4\xi-1}}}{1-c_2^2\kappa\xi A^{\frac{4\xi}{4\xi-1}}},
 \label{nt}
 \en
 where
 \be
 A=-e^{2c_1b\xi}+(4\xi-1)e^{bt}.
 \en
 In the derivation of (\ref{ps}), we have used $|\epsilon_{i}|<<1$ and we take the value of the scalar perturbation when reentering
 Hubble radius. COBE normalization constrains $ {\cal P}^{1/2}_{S}(k=aH)=1.9\times 10^{-5}$ \cite{cobe}. The perturbation freezes when it exits the Hubble radius. In other words, when the perturbations are outside the horizon, the perturbations are subject to a certain conservation law, that permit us to connect the COBE normalization value to the perturbation value when it exits Hubble radius.
 From (\ref{ns}) and (\ref{nt}) it is clear that both the scalar
 index and tensor index are not exactly scale invariant. So we reach the conclusion that a non-exact scale invariant
 spectrum can be generated in an exact de Sitter space.

  Though we get the exact expressions of $n_S$ and $n_T$, we hardly
 say anything on observational effects of our exact de Sitter, since there are several free constants in these expressions. However, in fact, we have
 natural choices for most of these parameters. For example, the conformal coupling case is the most important case of STG, thus we
 set $\xi=1/6$. From (\ref{phis}), $c_2$ must be non-zero to evade a trivial $\phi$, $c_1$ is determined by the initial condition
 of $\phi$.  Note that this initial condition does not inhabit at the big bang epoch. From (\ref{scale}), $t=0$ does not imply big bang. This solution
  must match to another solution with big bang if one requires there is a big bang in the history of the universe, and then makes a (tiny) time
  translation to recover our ordinary cosmic time. The parameter $b$ is the energy scale of inflation, which can be up to the Planck scale or down to
  TeV scale. To avoid the horizon problem and flatness problem, we take that universe expands during inflation by a factor 60 e-foldings. From (\ref{phis}), one sees
  \be
  c_2^2\kappa\leq 1,
  \label{c2}
  \en
  since $\phi$ should be at the Planck scale at most. Now we introduce $l, m, n$ to measure the
  energy scale of $c_1, c_2, b$,
  \be
  l=bc_1,~m=c_2\sqrt{\kappa},~n=b\sqrt{\kappa},
  \label{kmn}
  \en
  then (\ref{ns}), (\ref{nt}), and (\ref{ps}) reduce at $\eta=0.002{\rm Mpc}^{-1}$ to
  \be
  n_S-1=\frac{2[(6+9m^2)B^{1/3}-18B^{2/3}-54B+2]}{(18B^{2/3}+12B^{1/3}-3m^2+2)(3B^{1/3}+1)},
  \label{ns1}
  \en
   \be
  n_T=-\frac{6m^2}{(18B^{2/3}+12B^{1/3}-3m^2+2)(3B^{1/3}+1)},
  \label{nt1}
  \en
 and
 \be
 {\cal P}^{1/2}_{S}=\frac{n}{6\sqrt{2}m\pi}(1+3B^{1/3})\sqrt{18B^{2/3}+12B^{1/3}-3m^2+2}~,
 \label{ps1}
 \en
 where $B=e^{l}$.

 Now we make an analysis of a concrete model based on our solution. As we mentioned, the scale of $\phi$ should not
 be higher than Planck scale, thus we have (\ref{c2}), which yields $m\leq 1$. Here we take $m=0.01$. At the largest scale $\eta=0.002 {\rm Mpc}^{-1}$ or the wave length of
 perturbation $L=3000$Mpc, observation data from WMAP-7 implies $n_S-1=0.027$ \cite{wmap7}, which requires $l=-3.38$ from (\ref{ns1}). With this construction,
 $n_T=-3.9\times 10^{-5}$, and the amplitude of scalar perturbation becomes,
 \be
 {\cal P}^{1/2}_{S}=20.6n=1.9\times 10^{-5}.
 \en
 Then we derive the energy scale of the inflation $b$ by using (\ref{kmn}),
 \be
 b=0.92\times 10^{-6}\kappa^{-1/2}\sim 10^{12} {\rm Gev}.
 \en
 This is a reasonable scale for inflation. A significant prediction of this construction is that the tensor perturbation
  is strenuously depressed, since $r=2|n_T|=7.8\times 10^{-5}$, which is far below the lowerbound of any present probes. This depression of
  tensor mode is an interesting point which deserves to study a bit more. (\ref{c2}) imposes a rough restriction $c_2$. From (\ref{phisimple}),
  a refined upperbound of $c_2$ should be
   \be
  c_2^2\kappa\leq \frac{1}{9},
  \label{c2p}
  \en
  since $c_1<0$ to be consistent with the observed amplitude of the scalar perturbation. After a lengthy numerical analysis, we obtain $l<-3.24$ by using (\ref{ns1}).
  Substituting these results into (\ref{nt1}), we arrive at,
  \be
  |n_T|<0.042,
  \en
  by numerical method. Thus $r<0.084$. This is an explicit predication of this model, which is about an order below the ability of WMAP-7, whose data constrain
  $r<0.49$ (2$\sigma$ CL). Certainly, as we have seen, this model prepares enough space for a very weak tensor mode. But if observations display that
  the amplitude of tensor mode is stronger than 0.084 times of the scalar mode, this model will be unfavored, though we can select some unnatural parameters to save it.

\section{Conclusion~ and~ discussion}  This de Sitter ground state is important since both the early inflation and the late time acceleration can be regarded as fluctuations on
  a de Sitter background. We first find a de Sitter solution of STG solution. We expect that this solution may shed some light on the de Sitter vacua in string/M. The potential of the non-minimal
 coupling scalar takes a multinomial power-law form. The potential reduces to a simple form, which combines the constant and $\phi^{4}$ term, in the conformal coupling case. In this solution, the scalar rolls down from a power-law
 potential though the geometry is exactly a de Sitter. The stability property of this solution is investigated. We find that it is stable in the weak coupling limit. In Einstein frame, this solution can describe a universe which decelerates in the early time and accelerates in the late time.

 Based on this solution, we explore its inflationary theory in frame of STG. We studied the primordial
 quantum perturbation around the adiabatic vacuum in our solution. We get the spectrum of the scalar mode and tensor mode of the
 quantum perturbation. In the case of a conformal coupling scalar, the model with the parameter set normalized by the amplitude of the scalar perturbation predicts
 a strongly depressed tensor mode. We expect the future observation, especially observation from CMB and other primordial gravity wave probes, can improve this model.
 A problem to be investigated is the graceful exit problem. The present solution describes an exact de Sitter solution, which implies that the inflation phase never
 exits. There are two roads to solve this problem. First, we can introduce another term in the (\ref{potential}). In this case the space will be no longer an exact de Sitter space,
 thus the universe may exit the inflation phase at some epoch. Second, we can glue this solution to the other solution which can exit inflation. Under this situation, there are two
 stages of cosmic evolutions. At the first stage, all of our investigations still work. To find a proper solution to glue with our present solution is the future work.

{\bf Acknowledgments}
 We are grateful to Prof M Sasaki for his several useful suggestions. This work is supported by National Education Foundation of China under grant No. 200931271104,
 Shanghai Municipal Pujiang grant No. 10PJ1408100, and National Natural Science Foundation of China under Grant No. 11075106.

\end{document}